\documentclass[10pt, conference, letterpaper]{IEEEtran}
\usepackage[letterpaper, left = 0.68in, right =0.68in, top = 0.7in, bottom = 1in]{geometry}
\def\arxiv{1}
\def\arxivdisclaimer{1} 
\def\isshorter{0}
\def\review{0} 
\usepackage[noadjust]{cite}
\if\review1
\usepackage{todonotes}
\else
\usepackage[disable]{todonotes}
\fi
\usepackage[acronym]{glossaries}
\usepackage{amsthm}
\usepackage{amsmath, amssymb}
\usepackage{balance}

\usepackage{svg} 
\usepackage{graphicx} 
\usepackage{pgfplots} 
\usetikzlibrary{spy} 
\usetikzlibrary{arrows.meta} 
\usepackage{xfrac} 
\pgfplotsset{compat=newest} 
\pgfplotsset{
    /pgfplots/layers/Bowpark/.define layer set={
        axis background,axis grid,main,axis ticks,axis lines,axis tick labels,
        axis descriptions,axis foreground
    }{/pgfplots/layers/standard},
    colormap={jet}{
        rgb255(0cm)=(0,0,128);
        rgb255(1cm)=(0,0,255);
        rgb255(3cm)=(0,255,255);
        rgb255(5cm)=(255,255,0);
        rgb255(7cm)=(255,0,0);
        rgb255(8cm)=(128,0,0)
    }
} 

\if\arxiv0
\else
\usepackage{hyperref} 

\glsdisablehyper
\fi
\usepackage{cleveref}
\usepackage{subcaption}
\usepackage{makecell} 
\usepackage{lipsum} 

\usepackage{textcase}
\usepackage[tablename=TABLE,font=small]{caption}
\DeclareCaptionTextFormat{up}{\MakeTextUppercase{#1}}
\newcommand{\wrt}{w.\,r.\,t.\ }

\newcommand\blfootnote[1]{%
  \begingroup
  \renewcommand\thefootnote{}\footnote{#1}%
  \addtocounter{footnote}{-1}%
  \endgroup
}

\newtheorem{remark}{Remark}
\newtheorem*{remark*}{Remark}

\begin{document}
\title{Optimal Azimuth Sampling and Interpolation for Bistatic ISAC Setups}

\author{

\IEEEauthorblockN{
        Alexander~Felix\IEEEauthorrefmark{1}\IEEEauthorrefmark{2},
        Silvio Mandelli\IEEEauthorrefmark{1},
        Marcus~Henninger\IEEEauthorrefmark{1}, 
        and Stephan ten Brink\IEEEauthorrefmark{2} 
        }
        
	\IEEEauthorblockA{
	\IEEEauthorrefmark{1}Nokia Bell Labs Stuttgart, 70469 Stuttgart, Germany \\
	\IEEEauthorrefmark{2}Institute of Telecommunications, University of Stuttgart, 70569 Stuttgart, Germany \\
	E-mail: alexander.felix@nokia.com}        
\thanks{TBD: Further notes.}}


%


\maketitle

\newacronym{2D}{2D}{two-dimensional}
\newacronym{5G}{5G}{fifth generation}
\newacronym{6G}{6G}{sixth generation}
\newacronym{awgn}{AWGN}{additive white Gaussian noise}
\newacronym{cfar}{CFAR}{constant false alarm rate}
\newacronym{dft}{DFT}{discrete Fourier transform}
\newacronym{isac}{ISAC}{Integrated Sensing and Communications}
\newacronym[plural={MRAs}]{mra}{MRA}{minimum redundancy array}
\newacronym{lmf}{LMF}{Location Management Function}
\newacronym{mf}{MF}{management function}
\newacronym{naf}{NAF}{normalized angular frequency}
\newacronym{NRPPa}{NRPPa}{NR Positioning Protocol A}
\newacronym{poc}{PoC}{proof of concept}
\newacronym[plural={PSFs}]{psf}{PSF}{point spread function}
\newacronym[plural={RUs}]{ru}{RU}{radio unit}
\newacronym{rx}{RX}{receiver}
\newacronym{rmse}{RMSE}{root mean squared error}
\newacronym{sar}{SAR}{synthetic-aperture radar}
\newacronym{sara}{SARA}{Sampling and Reconstructing Angular Domains with Uniform Arrays}
\newacronym{semf}{SeMF}{Sensing Management Function}
\newacronym{tx}{TX}{transmitter}
\newacronym[plural={TRPs}]{trp}{TRP}{transmission reception point}
\newacronym[plural={ULAs}]{ula}{ULA}{uniform linear array}
\newacronym[plural={URAs}]{ura}{URA}{uniform rectangular array}

\if\isshorter1
\begin{abstract}

A key challenge in future 6G \acrfull{isac} networks is to define the angular operations of \acrlong{tx} and \acrlong{rx}, i.e., the sampling task of the angular domains, to acquire information about the environment. 
In this work we extend previous analysis for optimal angular sampling of monostatic setups to two-dimensional bistatic deployments, that are as important as the former in future \acrshort{isac} cellular scenarios.

Our approach 
overcomes the limitations of suboptimal prior art sampling and interpolation techniques, such as spline interpolation. We demonstrate that separating azimuth operations of the two transmit and receive arrays is optimal to sample the angular domain in an array-specific \gls{naf}. This allows us to derive a loss-less reconstruction of the angular domain, enabling a more efficient and accurate sampling strategy for bistatic sensing applications compared to legacy approaches.
As demonstrated by different Monte Carlo experiments, our approach enables future bistatic \acrshort{isac} deployments with {better performance} compared to the other suboptimal solutions.




\end{abstract}

\else

\fi

\if\arxivdisclaimer1
\blfootnote{This work has been submitted to the IEEE for possible publication. Copyright may be transferred without notice, after which this version may no longer be accessible.}
\vspace{-0.50cm}
\else
\vspace{0.2cm}
\begin{IEEEkeywords}
Bistatic radar, Bistatic sampling and interpolation, Integrated Sensing and Communications (ISAC), Imaging.
\end{IEEEkeywords}
\fi

\IEEEpeerreviewmaketitle

\if\isshorter1
\glsresetall
\section{Introduction}

As part of the new 6G feature set, \gls{isac} aims to expand the capabilities of cellular networks beyond communications by adding radar-like functionalities. Consequently, a wide range of potential \gls{isac} applications and deployments for 6G cellular networks is being explored~\cite{3GPP_TR_22837}. Ideally, \gls{isac} services could be enabled with minimal impact on current deployments~\cite{wild2021joint}. One type of deployment are monostatic array setups, which require either full-duplex capable arrays or the supplementary installation of a co-located receiver array~\cite{felixAntennaArrayDesign2024a}. However, neither of these configurations has yet been commercialized in a field setting, but only showcased in \glspl{poc}~\cite{wild2023integrated}. 
In contrast, bistatic deployments of two displaced arrays with common coverage areas can be found within current cellular networks. 

In radar literature, bistatic setups represent only a minor proportion of the deployed systems~\cite{richardsFundamentalsRadarSignal2014}. Furthermore, even within bistatic radar research, there is notable emphasis on the areas of \gls{sar}~\cite{skolnikIntroductionRadarSystems2002}, aviation or nautical radar~\cite{sitCharacterizingEvaporationDucts2019}, and passive radar~\cite{panFeasibilityStudyPassive2019}. Therefore, bistatic radar with fixed \gls{tx} and \gls{rx}, as in communications deployments, is not a prominent focus of previous investigations.

For the angular sampling task with confined acquisition capabilities, e.g., considering analog or hybrid beamforming, a set of angular samples must be chosen that optimally provides a complete representation of the angular domain. In the monostatic scenario, this challenge has been addressed by ~\cite{rajamaki2019analog, rajamaki2020hybrid}, with the \gls{dft} based \textit{\gls{sara}} approach~\cite{mandelli2022sampling} subsequently providing an optimal solution.

In bistatic scenarios, the angular steering operations of the separated arrays are independent, increasing the complexity of the angular sampling task compared to monostatic deployments~\cite{skolnikIntroductionRadarSystems2002}. Given the increased sampling space of a double set of azimuth and elevation angles, minimal yet optimal sampling schemes are even more important for bistatic deployments with confined angular acquisition capabilities.

If one wants to sense an environment with a bistatic setup, one has to define which angles are scanned by the \gls{tx}, thus defining the set of AoDs, and for each of them, which angles the \gls{rx} scans at the same time. Determining these two sets is the problem addressed in this work, which is particularly relevant for systems of confined angular capabilities, e.g., systems with analog and hybrid beamforming structures.




Our approach introduces an optimal sampling methodology for the \gls{2D} azimuth domain of such confined bistatic setups. 
Our contribution centers on:
\begin{itemize}
    \item Transferring the \gls{dft} based optimal sampling and loss-less interpolation method \gls{sara}~\cite{mandelli2022sampling}, defined for monostatic setups, to bistatic deployments on a per array basis; enabled by
    \item Establishing decomposed sets of azimuth operations by the \gls{tx} and \gls{rx} arrays in their respective \gls{naf} domains needed for the optimal \textit{angular sampling task}; facilitating the implementation of bistatic \gls{isac} deployments with minimal overhead for use cases necessitating solely azimuthal discrimination; validated by
    \item Demonstrating \textit{better performance} along multiple metrics than suboptimal prior solutions in a variety of Monte Carlo experiments.
\end{itemize}





\section{System Model}

We consider two displaced \glspl{ula} for our bistatic setup operating within the azimuth component of the angular domain. The spatial separation of the arrays in bistatic setups is referred to as the bistatic baseline. As shown in Fig.~\ref{fig:bistatic_angles}, the azimuth angle is defined between the corresponding direction vector and the boresight of an array, with positive values for clockwise rotation.

\subsection{Bistatic Setup of Displaced Uniform Linear Arrays}

Without loss of generality and as illustrated in Fig.~\ref{fig:bistatic_angles}, we assume that the \gls{tx} \gls{ula} is located at coordinates $(-c, 0)$ and the \gls{rx} \gls{ula} at coordinates $(c, b)$. The boresight of the \gls{tx} is defined such that it is oriented in the positive direction of the y-axis, while the boresight of the \gls{rx} is arbitrarily oriented.

Henceforth, parameters associated with either array type are denoted by the superscript symbol (·). The specific cases of the \gls{tx} array and the \gls{rx} array are denoted by superscripts (t) and (r), respectively.

The number of elements of each \gls{ula} is denoted as $N^{(\cdot)}$. Further, the individual element positions $\mathbf{p}_i^{(\cdot)}=[x_i,0]^\intercal$ are aggregated per array in a set as $\mathcal{A}^{(\cdot)} = \left\{\mathbf{p}_i^{(\cdot)}  \Big| i \leq N^{(\cdot)}, i \in \mathbb{N} \right\}$.



\begin{figure}
  \centering

    \resizebox{0.99\columnwidth}{!}{
    \definecolor{darkorangegnu}{RGB}{255,127,14}
\definecolor{steelbluegnu}{RGB}{31,119,180}
\def \dcolor{black}

\def \globalscale {1.000000}
\begin{tikzpicture}[y=1pt, x=1pt, inner sep=0pt, outer sep=0pt]

    \def \axiscolor{black!40}
    \def \centerX{150.0}
    \def \centerY{70.0}
    
    \draw[color=\axiscolor, -{Stealth[scale=1.3,scale width=1.5, color=\axiscolor]}] (30.0, \centerY) -- (270.0, \centerY);
    \node[text=\axiscolor,anchor=south] (text8) at (260.0, \centerY-7.0){\footnotesize{x}}; 
    
    \draw[color=\axiscolor, -{Stealth[scale=1.3,scale width=1.5, color=\axiscolor]}] (\centerX, 0.0) -- (\centerX, 200.0);
    \node[text=\axiscolor,anchor=south] (text9) at (\centerX-4.0, 183.0){\footnotesize{y}}; 
    


    \def \scanTargetX {170.0}
    \def \scanTargetY {170.0}
    \draw[color=\axiscolor, dotted, ] (150.0, \scanTargetY) -- (\scanTargetX, \scanTargetY);
    \draw[color=\axiscolor, ] (148.0, \scanTargetY) -- (152, \scanTargetY);
    \node[text=\axiscolor,anchor=south] (text8) at (143.0, \scanTargetY-3){\footnotesize{$y_s$}};

    \draw[color=\axiscolor, dotted, ] (\scanTargetX, 70.0) -- (\scanTargetX, \scanTargetY);
    \draw[color=\axiscolor, ] (\scanTargetX, 68.0) -- (\scanTargetX, 72.0);
    \node[text=\axiscolor,anchor=south] (text8) at (\scanTargetX-1, \centerY-10.0){\footnotesize{$x_s$}};

    \node[text=teal,anchor=south west,align=center, font=\scriptsize] (text14) at (\scanTargetX-12.0, \scanTargetY+3.0){Scan\\ direction};


    \def \arrayCommX {80.0}
    \def \arrayCommY {70.0}
    \def \arraycolor {cyan}
    \def \arraytext {Transmitter}
    
    \node[text=black,anchor=south,align=center, font=\scriptsize] (text25) at (\arrayCommX, \arrayCommY-45){\textcolor{\arraycolor}{\arraytext}\\\textcolor{\arraycolor}{Array}};

    \node at (\arrayCommX, \arrayCommY-14) {
    \begin{tikzpicture}[scale=0.7, transform shape]
        \path[draw=\arraycolor,miter limit=10.0] (\arrayCommX+1.2, \arrayCommY-40.4) -- (\arrayCommX+1.2, \arrayCommY-9.3);
        \path[draw=\arraycolor,miter limit=10.0] (\arrayCommX-0.3, \arrayCommY-40.4) -- (\arrayCommX-0.3, \arrayCommY-9.3);
        \path[draw=\arraycolor,fill=white,rounded corners=0.6pt] (\arrayCommX-1.4, \arrayCommY-19.4) rectangle 
        (\arrayCommX+2.3, \arrayCommY-25.4);
        \path[draw=\arraycolor,fill=white,miter limit=10.0] (\arrayCommX-3.7, \arrayCommY-14.4) -- (\arrayCommX-11.6, \arrayCommY-14.4)
        .. controls (\arrayCommX-16.6, \arrayCommY-14.4) and (\arrayCommX-17.9, \arrayCommY-12.1) .. (\arrayCommX-15.6, \arrayCommY-7.7) -- (\arrayCommX-13.6, \arrayCommY-3.8)
        .. controls (\arrayCommX-12.2, \arrayCommY-1.2) and (\arrayCommX-9.0, \arrayCommY) .. (\arrayCommX-4.0, \arrayCommY) -- (\arrayCommX+11.7, \arrayCommY)
        .. controls (\arrayCommX+16.7, \arrayCommY) and (\arrayCommX+18.1, \arrayCommY-2.2) .. (\arrayCommX+15.8, \arrayCommY-6.6) -- (\arrayCommX+13.7, \arrayCommY-10.5)
        .. controls (\arrayCommX+12.4, \arrayCommY-13.1) and (\arrayCommX+9.2, \arrayCommY-14.4) .. (\arrayCommX+4.2, \arrayCommY-14.4) -- cycle;
    \end{tikzpicture}
    };

    \path[draw=black,fill=black] (\arrayCommX, \arrayCommY) ellipse (1pt and 1pt);  

    \draw[\arraycolor, thick] (\arrayCommX, \arrayCommY+40) arc (90:48:40);
    \node[\arraycolor] at (\arrayCommX+10, \arrayCommY+29) {\footnotesize{$\theta^\text{(t)}$}};
    \draw[\arraycolor, thick] (\scanTargetX, \scanTargetY-40) arc (270:228:40);
    \node[\arraycolor] at (\scanTargetX-10, \scanTargetY-29) {\footnotesize{$\theta^\text{(t)}$}};

    \draw[-{Stealth[scale=1.3,scale width=1.5,color=teal]}, thick, color=teal] (\arrayCommX, \arrayCommY) -- (\scanTargetX, \scanTargetY); 
    \node[text=teal,anchor=south] (text20) at (\arrayCommX+50.0, 109.5){\footnotesize{\(\mathbf{u}^{(\text{t})}\)}};
    \draw[\axiscolor, -{Stealth[scale=1.3,scale width=1.5,color=\axiscolor]}] (\arrayCommX, \arrayCommY) -- (\arrayCommX, 180.0); 
    \node[text=\axiscolor,anchor=south, align=center, font=\scriptsize] (text24) at (\arrayCommX-15, 165.0){TX\\boresight};

    \draw[color=\axiscolor, dotted, ] (\arrayCommX, \arrayCommY) -- (\arrayCommX, \centerY);
    \draw[color=\axiscolor, ] (\arrayCommX, \centerY+2.0) -- (\arrayCommX, \centerY-2.0);
    \node[text=\axiscolor,anchor=south] (text8) at (\arrayCommX, \centerY-7){\footnotesize{$-c$}};

    \def \arraySensX {220.0}
    \def \arraySensY {50.0}
    \def \arraycolor {darkorangegnu}
    \def \arraytext {Receiver}
    
    \node[text=black,anchor=south,align=center, font=\scriptsize] (text25) at (\arraySensX, \arraySensY-45){\textcolor{\arraycolor}{\arraytext}\\\textcolor{\arraycolor}{Array}};

    \node at (\arraySensX, \arraySensY-14) {
    \begin{tikzpicture}[scale=0.7, transform shape]
        \path[draw=\arraycolor,miter limit=10.0] (\arraySensX+1.2, \arraySensY-40.4) -- (\arraySensX+1.2, \arraySensY-9.3);
        \path[draw=\arraycolor,miter limit=10.0] (\arraySensX-0.3, \arraySensY-40.4) -- (\arraySensX-0.3, \arraySensY-9.3);
        \path[draw=\arraycolor,fill=white,rounded corners=0.6pt] (\arraySensX-1.4, \arraySensY-19.4) rectangle 
        (\arraySensX+2.3, \arraySensY-25.4);
        \path[draw=\arraycolor,fill=white,miter limit=10.0] (\arraySensX-3.7, \arraySensY-14.4) -- (\arraySensX-11.6, \arraySensY-14.4)
        .. controls (\arraySensX-16.6, \arraySensY-14.4) and (\arraySensX-17.9, \arraySensY-12.1) .. (\arraySensX-15.6, \arraySensY-7.7) -- (\arraySensX-13.6, \arraySensY-3.8)
        .. controls (\arraySensX-12.2, \arraySensY-1.2) and (\arraySensX-9.0, \arraySensY) .. (\arraySensX-4.0, \arraySensY) -- (\arraySensX+11.7, \arraySensY)
        .. controls (\arraySensX+16.7, \arraySensY) and (\arraySensX+18.1, \arraySensY-2.2) .. (\arraySensX+15.8, \arraySensY-6.6) -- (\arraySensX+13.7, \arraySensY-10.5)
        .. controls (\arraySensX+12.4, \arraySensY-13.1) and (\arraySensX+9.2, \arraySensY-14.4) .. (\arraySensX+4.2, \arraySensY-14.4) -- cycle;
    \end{tikzpicture}
    };

    \path[draw=black,fill=black] (\arraySensX, \arraySensY) ellipse (1pt and 1pt);  

    \draw[\arraycolor, thick] (\arraySensX, \arraySensY+40) arc (90:112:40);
    \node[\arraycolor] at (\arraySensX-6, \arraySensY+33) {\footnotesize{$\theta^\text{(r)}$}};
    \draw[\arraycolor, thick] (\scanTargetX, \scanTargetY-40) arc (270:292:40);
    \node[\arraycolor] at (\scanTargetX+7, \scanTargetY-33) {\footnotesize{$\theta^\text{(r)}$}};

    \draw[color=\axiscolor, dotted, ] (\centerX, \arraySensY) -- (\arraySensX, \arraySensY);
    \draw[color=\axiscolor, ] (\centerX-2.0, \arraySensY) -- (\centerX+2.0, \arraySensY);
    \node[text=\axiscolor,anchor=south] (text8) at (\centerX-8.0, \arraySensY-3){\footnotesize{$b$}};

    \draw[-{Stealth[scale=1.3,scale width=1.5,color=teal]}, thick, color=teal] (\arraySensX, \arraySensY) -- (\scanTargetX, \scanTargetY); 
    \node[text=teal,anchor=south] (text20) at (\arraySensX-15.0, 109.5){\footnotesize{\(\mathbf{u}^{(\text{r})}\)}};
    \draw[\axiscolor, -{Stealth[scale=1.3,scale width=1.5,color=\axiscolor]}] (\arraySensX, \arraySensY) -- (\arraySensX, 180.0); 
    \node[text=\axiscolor,anchor=south, align=center, font=\scriptsize] (text24) at (\arraySensX+15, 165.0){RX\\boresight};

    \draw[color=\axiscolor, dotted, ] (\arraySensX, \arraySensY) -- (\arraySensX, \centerY);
    \draw[color=\axiscolor, ] (\arraySensX-2.0, \centerY+2.0) -- (\arraySensX+2.0, \centerY-2.0);
    \node[text=\axiscolor,anchor=south] (text8) at (\arraySensX+4, \centerY-7){\footnotesize{$c$}};

    \def \targetX {125.0}
    \def \targetY {150.0}

    \path[draw=black,fill=violet] (\targetX, \targetY) ellipse (3pt and 3pt); 
    
    \draw[-{Stealth[scale=1.3,scale width=1.5,color=violet]}, color=violet] (\arrayCommX, \arrayCommY) -- (\targetX-2.6, \targetY-1.3); 
    \node[text=violet,anchor=south, font=\footnotesize] (text20) at ({\arrayCommX+(\targetX-\arrayCommX)/2-7.0}, {\arrayCommY+(\targetY-\arrayCommY)/2+1.0}){\(\mathbf{s}^{(\text{t})}\)};
    \draw[-{Stealth[scale=1.3,scale width=1.5,color=violet]}, color=violet] (\arraySensX, \arraySensY) -- (\targetX+2.4, \targetY-1.3); 
    \node[text=violet,anchor=south, font=\footnotesize] (text20) at ({\targetX+(\arraySensX-\targetX)/2-11.0}, {\arrayCommY+(\targetY-\arrayCommY)/2-12.0}){\(\mathbf{s}^{(\text{r})}\)};

    \node[text=violet,anchor=south west,align=center, font=\scriptsize] (text14) at (\targetX-16.0, \targetY+5.0){Scatterer\\ direction};


\end{tikzpicture}}
    \caption{Top-down view of a bistatic setup with distinct scan and scatterer directions, exemplarily showing the azimuth scan angles.}
    \label{fig:bistatic_angles}
    \vspace{-4mm}
\end{figure}
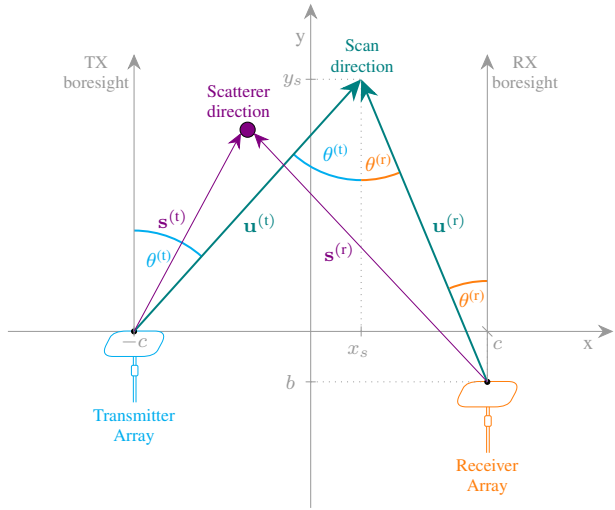

\subsection{Bistatic Angular Sampling Task}
The objective is to obtain a complete \gls{2D} image of the environment. It is impractical to sample all infinite \gls{tx}-\gls{rx} angle pairs. Consequently, the angular sampling task is defined as the problem of ``identifying the set of pairs given by the simultaneously sampled \gls{tx} and \gls{rx} angles''. The set of angular operations for the combined bistatic setup is denoted as $\Upsilon$, while the set for an individual array is denoted as $\mathcal{U}^{(\cdot)}$. A limitation on the number of simultaneously sampled pairs in this approach is fully consistent with analog and hybrid structures available in commercial use.

For an optimal solution for monostatic setups, we recall that the angular sampling problem was solved by~\cite{mandelli2022sampling} using the coarray theory introduced in~\cite{hoctor1990unifying}.

Since our methodology is also concerned with the optimal sampling and reconstruction of the angular response, our model excludes path loss from consideration, as this phenomenon is associated with the link budget rather than beamforming operations. To limit the analysis to the fundamental challenge of angular sampling, we assume synchronization between the two bistatic arrays.

First, we define an arbitrary direction vector as
\begin{equation}
\mathbf{v}^{(\cdot)}= \left[\sin\left(\theta^{(\cdot)}\right), {\cos\left(\theta^{(\cdot)}\right)} \right]^\intercal \; ,
\label{eq:DirectionVector}
\end{equation}
where $\theta$ is the considered azimuth angle. 
Further, a pair of scan directions of the bistatic setup is defined as $\left(\mathbf{u}^\text{(t)}, \mathbf{u}^\text{(r)}\right)$, while a pair of directions for a single scatterer is defined as $\left(\mathbf{s}^\text{(t)}, \mathbf{s}^\text{(r)}\right)$. The set of all scatterer direction pairs of a given scenario is defined as $\Psi$.

The narrowband imaging task is to estimate the deterministic amplitude function $a\left(\mathbf{s}^\text{(t)}, \mathbf{s}^\text{(r)}\right)$ of a coherent imaging signal with wavelength $\lambda_0$  at center frequency $\omega_0$ for a pair of \gls{tx}-\gls{rx} scan directions.
By applying linear beamforming to scan directions $\mathbf{u}^\text{(t)}$ and $\mathbf{u}^\text{(r)}$~\cite{hoctor1990unifying}, respectively, the coherent beamformed signal of a bistatic setup can be expressed as
\begin{align}
\thinmuskip=0.0mu
\medmuskip=0.1mu
\thickmuskip=0.5mu
\hat{a}\left(\mathbf{u}^\text{(t)}, \mathbf{u}^\text{(r)}\right) 
&= \sum_{n=1}^{N^\text{(t)}} \sum_{m=1}^{N^\text{(r)}} w_{n}^{(\text{t})}  w_{m}^{(\text{r})}  e^{-j k_0 \left( \mathbf{u^\text{(t)}}^\intercal \mathbf{p}_n^\text{(t)} + \mathbf{u^\text{(r)}}^\intercal \mathbf{p}_m^\text{(r)} \right)} \nonumber \\
& \;\;\; \cdot \iint\limits_\Psi a\left(\mathbf{s}^\text{(t)}, \mathbf{s}^\text{(r)}\right) e^{j k_0 \left(\mathbf{s^\text{(t)}}^\intercal \mathbf{p}_n^\text{(t)} + \mathbf{s^\text{(r)}}^\intercal \mathbf{p}_m^\text{(r)} \right)} d\mathbf{s}^\text{(t)} d\mathbf{s}^\text{(r)}\;,
\label{eq:EstimatedImageBi}
\end{align}
where $k_0 = 2\pi/\lambda_0$ is the wavenumber and $w^{(\cdot)}$ is the beamforming coefficient of a single array element. 
\begin{remark}
To minimize the width of the beamformer's main lobe, and thus maximize the resolution capabilities, in the following equal amplitudes are chosen for each element. However, our findings apply, for any distribution of beamforming coefficients, as in~\cite{mandelli2022sampling}. 
\end{remark}

In the monostatic case, $\mathbf{u}^{(t)} = \mathbf{u}^{(r)}$, the phase term of Eq.~\eqref{eq:EstimatedImageBi} simplifies, leading to the coarray formulations found in~\cite{hoctor1990unifying}. 
However, it is evident that this joint aggregation is not feasible for the bistatic case due to $\mathbf{u}^{(t)} \neq \mathbf{u}^{(r)}$. 
Accordingly, this distinction poses the problem of designing operations to be applied in bistatic sensing scenarios.

\section{Bistatic Azimuth Sampling and Interpolation}


We recall the bistatic angular sampling task, which consists in defining a set of \gls{tx}-\gls{rx} angle pairs to be scanned by our \glspl{ula} setup. For a single pair of angles consisting of an arbitrary combination of azimuth scan angles $\theta_s^{(\cdot)}$, the combined phase term of the beamforming operations is
\begin{align}
\thinmuskip=1.5mu
\medmuskip=1.5mu
\thickmuskip=2mu
\varphi\left(\theta_s^\text{(t)},\theta_s^\text{(r)}\right) &= \sum_{n=1}^{N^\text{(t)}}\sum_{m=1}^{N^\text{(r)}} e^{-j k_0 \left( \mathbf{u^\text{(t)}}^\intercal \mathbf{p}_n^\text{(t)} + \mathbf{u^\text{(r)}}^\intercal \mathbf{p}_m^\text{(r)} \right)} \nonumber \\
&= \sum_{n=1}^{N^\text{(t)}} e^{-jk_0\sin{\left(\theta^{\text{(t)}}\right)}x_n} \sum_{m=1}^{N^\text{(r)}} e^{-jk_0\sin{\left(\theta^{\text{(r)}}\right)}x_m} \nonumber \\
&= \varphi^\text{(t)}\left(\theta_s^\text{(t)}\right) \cdot \varphi^\text{(r)}\left(\theta_s^\text{(r)}\right) \;,
\label{eq:BistaticPhaseTerm}
\end{align}
which provides a clear separation of the decomposed phase terms for each array.

By defining the aggregated scatterer effects of the scenario as amplitude coefficients $a_{n}^{(\text{t})}$ at the $n$-th \gls{tx} element and, respectively, $a_{m}^{(\text{r})}$ at the $m$-th \gls{rx} element,
we can reformulate \eqref{eq:EstimatedImageBi} as
\begin{align}
\thinmuskip=1.5mu
\medmuskip=1.5mu
\thickmuskip=2mu
\hat{a}\left(\theta_s^\text{(t)},\theta_s^\text{(r)}\right) = 
&\sum_{n=1}^{N^\text{(t)}} a_{n}^{(\text{t})} w_{n}^{(\text{t})} e^{-jk_0\sin{\left(\theta^{\text{(t)}}\right)}x_n} \nonumber \\
&\cdot \sum_{m=1}^{N^\text{(r)}} a_{m}^{(\text{r})} w_{m}^{(\text{r})} e^{-jk_0\sin{\left(\theta^{\text{(r)}}\right)}x_m} \;,
\label{eq:EstimateBiBeamform}
\end{align}
demonstrating for the individual phase terms of the arrays the well-known Fourier duality between the aperture plane defined by the element positions $\mathbf{p}_i$ and the radar source direction $\mathbf{u}^{(\cdot)}$ depending on the scan angle $\theta_s^{(\cdot)}$.

\subsection{Independent Angular Operation in Orthonormal Base}
For the completion of the angular sampling task, a set of \gls{tx}-\gls{rx} angle pairs, $\Upsilon$, has to be scanned. 
Therefore, we are interested in the changes of angular directions from one angle pair of interest to the next.
The separation of the phase terms and thus the angular operations in \eqref{eq:BistaticPhaseTerm} and \eqref{eq:EstimateBiBeamform} show that each arrays beamforming operation is based on individual orthonormal basis, one depending only on the \gls{tx}, the other only on the \gls{rx}. Therefore, the bistatic angular sampling task can be solved by solving the sampling task separately for the \gls{tx} and \gls{rx}, giving two $1$D problems.

For a single \gls{ula}, we utilize the Fourier duality to sample the remaining $1$D angular domain as suggested in~\cite{mandelli2022sampling}, that in case of element spacing by half the wavelength coincides with DFT beamforming. This corresponds to taking $N^{(\cdot)}$ samples for each array resulting in the sets $\mathcal{U}^{(\text{t})}$ and $\mathcal{U}^{(\text{r})}$. Finally, computing the Cartesian product of the two sets gives the combined set $\Upsilon=\mathcal{U}^{(\text{t})} \times \mathcal{U}^{(\text{r})}$, which completes the bistatic angular sampling task.


The Cartesian product of the two sampling sets can also be conceptualized as an $N^{(\text{t})} \times N^{(\text{r})}$ sampling grid. 
Thus, an iterative approach of sweeping the scan directions of one array while fixating the scan direction of the other array, i.e., completing the grid row-by-row or column-by-column. 

The selection of $N^{(\cdot)}$ uniformly spaced samples within the angular domain of each array combined with the uniformly spaced array elements facilitates the implementation of \gls{dft}-based optimal interpolation. 

This solution has the same properties of the derivations in~\cite{mandelli2022sampling}, thus ensuring loss-less reconstruction with minimal amount of angular samples $N^{(\text{t})} N^{(\text{r})}$.
In contrast to our approach, the suboptimal prior art approaches, such as spline interpolation, are only capable of approximating the angular response of the system.

\subsection{Translation of 2D NAF to Cartesian}

For a variety of applications, it is necessary to translate the the desired upsampled angular acquisition from an angular-based system into the global Cartesian coordinate system. 

From our setup in Fig.~~\ref{fig:bistatic_angles} and with trigonometric reformulations, we can extract the \gls{2D} Cartesian coordinates of any scan point $\mathbf{p}_s$ as
\begin{equation}
    \thinmuskip=0.5mu
    \medmuskip=0.5mu
    \thickmuskip=1.5mu
    \mathbf{p}_s= \begin{bmatrix}
       \dfrac{2c-b\tan\left({\theta}^\text{(r)}\right)}{\tan\left(\theta^\text{(t)}\right)-\tan\left({\theta}^\text{(r)}\right)} \\[6mm]
       \dfrac{c\left(\tan\left(\theta^\text{(t)}\right)+\tan\left({\theta}^\text{(r)}\right)\right) - b \tan\left(\theta^\text{(t)}\right) \tan\left({\theta}^\text{(r)}\right)}{\tan\left(\theta^\text{(t)}\right)-\tan\left({\theta}^\text{(r)}\right)}
    \end{bmatrix}^\intercal \; .
    \label{eq:Angle2Cartesian}
\end{equation}

It is also possible to consider a rotation between the boresight of one array to the other. As a first step, taking the \gls{tx} boresight as reference, we substitute the azimuth angle of the rotated \gls{rx} array ${\theta}^{(\text{r})}$ in Eq.~\eqref{eq:Angle2Cartesian} with $\hat{\theta}^{(\text{r})}$. Then, the global \gls{rx} azimuth angle relevant for the translation is defined as $\hat{\theta}^{(\text{r})}={\theta}^{(\text{r})}+\tilde{\theta}^{(\text{r})}$. The azimuth angle operation, as previously introduced w.r.t. the \gls{rx}'s own boresight, remains denoted by ${\theta}^{(\text{r})}$.
The additional component $\tilde{\theta}^{(\text{r})}$ covers the relative rotation between the boresights of \gls{rx} and \gls{tx}.


\section{Simulation Studies}


\renewcommand{\arraystretch}{1.30} 
\begin{table}
    \centering
    \caption{Simulation parameters and assumptions}
    \label{tab:sim_param}
     \begin{tabular}{|c|c|}
        \hline
        \textbf{Parameter} & \textbf{Value / Description} \\
        \Xhline{3\arrayrulewidth}
        TX/RX array & \glspl{ula} with $N^{(\cdot)}=11$\\
        \hline
        TX/RX element spacing & $\lambda/2$ \\
        \hline
        Distance between TX and RX  & 12 m \\
        \hline
        Number of targets & 2 \\
        \hline
        Noise power \(\sigma^2\) & \(10\) dB \\
        \hline
        Target reflection coefficients & \(e^{j\Phi_s}\), where \(\Phi_s\backsim U\left[0,2\pi\right)\) \\
        \hline
        Desired $P^\text{(FA)}$ of CFAR & \(10^{-3}\) \\
        \hline
        Iterations per data point & \(10^{4}\) \\
        \hline
    \end{tabular}
\vspace{-4mm}
\end{table}

A two-part evaluation is employed to validate our methodology. Firstly, a qualitative evaluation with 2D-azimuth plots is presented, followed by a quantitative Monte Carlo evaluation of different performance metrics. The investigated bistatic scenarios follow our setup in Fig.~\ref{fig:bistatic_angles} with the parameters provided in Tab. \ref{tab:sim_param}. The core setup consists of two displaced \glspl{ula} spaced $12$~m along the x-axis. The \gls{tx}, on the left side, and the \gls{rx}, on the right side, are used to observe two ideal point scatterers, which are positioned at random locations in the area but with a fixed distance from each other in the \gls{2D} azimuth space. Depending on the scenario, further restrictions on the placement of the scatterers are enforced.

\subsection{Baseline}
As a baseline, we use the \gls{2D} spline interpolation in two variations (as implemented in the SciPy package v1.14.0~\cite{virtanenSciPyFundamentalAlgorithms2020}). 

As incorporated in our approach, the first interpolation baseline operates directly on the optimal \gls{naf} samples, henceforth referred to as \textit{NAF spline}, while the second interpolation baseline operates with uniformly spaced samples acquired in the radian domain, henceforth referred to as \textit{RAD spline}.

\subsection{2D Evaluation}

\def \intplotsource {rect}  
\def\genvspace{-1.0mm}

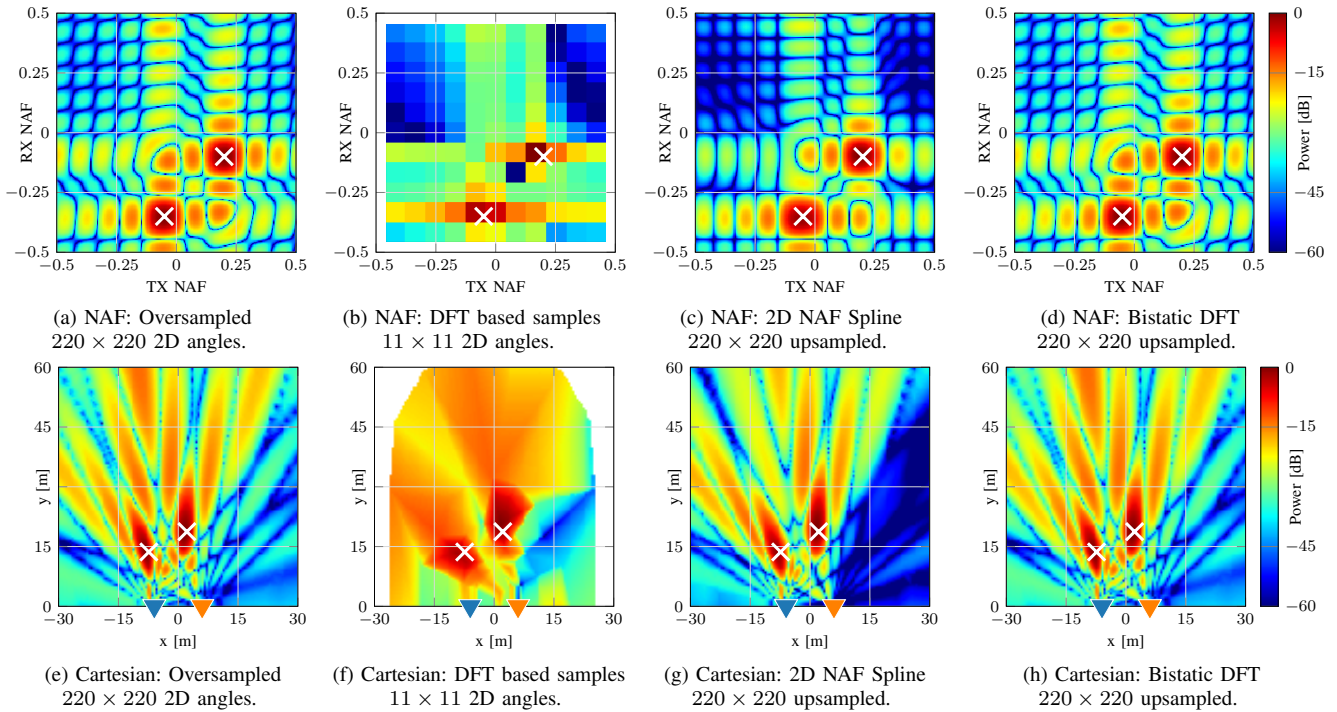
\begin{figure*}[ht]
  \def\subscale{.77}
  \centering
  \hspace{-3mm}
  \subfloat
  		[\gls{naf}: Oversampled \\$220\times220$ 2D angles. 
  		\label{fig:intensity_reference_oversampled}]
  		{
		\scalebox{\subscale}{\pgfplotsset{scaled y ticks=false}

\def \globalscale {1}
\begin{tikzpicture}[yscale=\globalscale,xscale=\globalscale]

    \begin{axis}[
        axis on top,
        height=0.315\textwidth,
        width=0.315\textwidth,
        grid=major,
        grid style={solid, black!15},
        enlargelimits=false,
        xmin=-0.5, xmax=0.5,
        xlabel near ticks,
        x label style={yshift=0.2em},
        ymin=-0.5, ymax=0.5,
        ytick={-0.5,-0.25,0,0.25,0.5},
        xtick={-0.5,-0.25,0,0.25,0.5},
        xticklabels={$-0.5$,$-0.25$,$0$,$0.25$,$0.5$},
        yticklabels={$-0.5$,$-0.25$,$0$,$0.25$,$0.5$},
        ylabel near ticks,
        xlabel={TX \gls{naf}},
        ylabel={RX \gls{naf}},
        y label style={yshift=-2em},
        label style={font=\footnotesize},
        tick label style={font=\footnotesize},
        legend style={font=\footnotesize},
        legend pos = north east,
        legend style=
        	{fill=white, 
        	fill opacity=0.4, 
        	draw opacity=1, 
        	text opacity=1, 
        	nodes={scale=1, transform shape}, 
            /tikz/every even column/.append style={column sep=0.1cm}
        	},
        colormap name=jet,
        ]

      \addplot[forget plot] graphics[
      xmin=-0.5, 
      xmax=0.5, 
      ymin=-0.5, 
      ymax=0.5,
      includegraphics={
        trim=1 0 0 1,
        clip,}
        ] {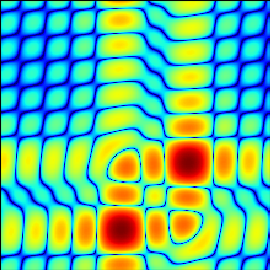};

    \addplot[mark=x, color=white, mark options={scale=2, line width=1.5pt}, mark size=3.0pt] coordinates {(-0.05, -0.35)};
    \addplot[mark=x, color=white, mark options={scale=2, line width=1.5pt}, mark size=3.0pt] coordinates {(0.2, -0.1)};

    \end{axis}

\end{tikzpicture}}
        \vspace{\genvspace}
		} 
		\hspace{-6mm}
  \subfloat
  		[\gls{naf}: \gls{dft} based samples \\$11\times11$ 2D angles. 
  		\label{fig:intensity_sara_sampled}]
  		{
		\scalebox{\subscale}{\pgfplotsset{scaled y ticks=false}

\def \globalscale {1}
\begin{tikzpicture}[yscale=\globalscale,xscale=\globalscale]

    \begin{axis}[
        axis on top,
        height=0.315\textwidth,
        width=0.315\textwidth,
        grid=major,
        grid style={solid, black!15},
        enlargelimits=false,
        xmin=-0.5, xmax=0.5,
        xtick={-0.6,-0.4,-0.2,0,0.2,0.4,0.6},
        xlabel near ticks,
        x label style={yshift=0.2em},
        ymin=-0.5, ymax=0.5,
        ytick={-0.5,-0.25,0,0.25,0.5},
        xtick={-0.5,-0.25,0,0.25,0.5},
        xticklabels={$-0.5$,$-0.25$,$0$,$0.25$,$0.5$},
        yticklabels={$-0.5$,$-0.25$,$0$,$0.25$,$0.5$},
        ylabel near ticks,
        xlabel={TX \gls{naf}},
        ylabel={RX \gls{naf}},
        y label style={yshift=-2em},
        label style={font=\footnotesize},
        tick label style={font=\footnotesize},
        legend style={font=\footnotesize},
        legend pos = north east,
        legend style=
        	{fill=white, 
        	fill opacity=0.4, 
        	draw opacity=1, 
        	text opacity=1, 
        	nodes={scale=1, transform shape}, 
            /tikz/every even column/.append style={column sep=0.1cm}
        	},
        ]

      \addplot[forget plot]graphics[
      xmin=-0.45454545, 
      xmax=0.46, 
      ymin=-0.45454545, 
      ymax=0.45454545,
      includegraphics={
        trim=1 0 1 1,
        clip,}
        ] {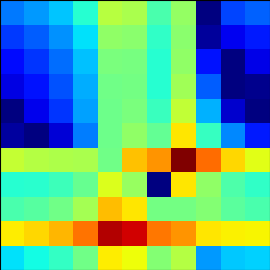};

    \addplot[mark=x, color=white, mark options={scale=2, line width=1.5pt}, mark size=3.0pt] coordinates {(-0.05, -0.35)};
    \addplot[mark=x, color=white, mark options={scale=2, line width=1.5pt}, mark size=3.0pt] coordinates {(0.2, -0.1)};

    \end{axis}

\end{tikzpicture}}
        \vspace{\genvspace}
		} 
		\hspace{-6mm}
  \subfloat
  		[\gls{naf}: 2D NAF Spline \\$220\times220$ upsampled.
  		\label{fig:intensity_baselineNAF}]
  		{
		\scalebox{\subscale}{\pgfplotsset{scaled y ticks=false}

\def \globalscale {1}
\begin{tikzpicture}[yscale=\globalscale,xscale=\globalscale]

    \begin{axis}[
        axis on top,
        height=0.315\textwidth,
        width=0.315\textwidth,
        grid=major,
        grid style={solid, black!15},
        enlargelimits=false,
        xmin=-0.5, xmax=0.5,
        xlabel near ticks,
        x label style={yshift=0.2em},
        ymin=-0.5, ymax=0.5,
        ytick={-0.5,-0.25,0,0.25,0.5},
        xtick={-0.5,-0.25,0,0.25,0.5},
        xticklabels={$-0.5$,$-0.25$,$0$,$0.25$,$0.5$},
        yticklabels={$-0.5$,$-0.25$,$0$,$0.25$,$0.5$},
        ylabel near ticks,
        xlabel={TX \gls{naf}},
        ylabel={RX \gls{naf}},
        y label style={yshift=-2em},
        label style={font=\footnotesize},
        tick label style={font=\footnotesize},
        legend style={font=\footnotesize},
        legend pos = north east,
        legend style=
        	{fill=white, 
        	fill opacity=0.4, 
        	draw opacity=1, 
        	text opacity=1, 
        	nodes={scale=1, transform shape}, 
            /tikz/every even column/.append style={column sep=0.1cm}
        	},
        colormap name=jet,
        ]

      \addplot[forget plot] graphics[
      xmin=-0.5, 
      xmax=0.5, 
      ymin=-0.5, 
      ymax=0.5,
      includegraphics={
        trim=1 0 0 1,
        clip,}
        ] {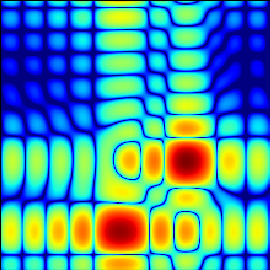};

    \addplot[mark=x, color=white, mark options={scale=2, line width=1.5pt}, mark size=3.0pt] coordinates {(-0.05, -0.35)};
    \addplot[mark=x, color=white, mark options={scale=2, line width=1.5pt}, mark size=3.0pt] coordinates {(0.2, -0.1)};

    \end{axis}

\end{tikzpicture}}
        \vspace{\genvspace}
		} 
		\hspace{-6mm} 
  \subfloat
  		[\gls{naf}: Bistatic \gls{dft} \\$220\times220$ upsampled.
  		\label{fig:intensity_sara_upsampled}]
  		{
		\scalebox{\subscale}{\pgfplotsset{scaled y ticks=false}

\def \globalscale {1}
\begin{tikzpicture}[yscale=\globalscale,xscale=\globalscale]

    \begin{axis}[
        axis on top,
        height=0.315\textwidth,
        width=0.315\textwidth,
        grid=major,
        grid style={solid, black!15},
        enlargelimits=false,
        xmin=-0.5, xmax=0.5,
        xtick={-0.6,-0.4,-0.2,0,0.2,0.4,0.6},
        xlabel near ticks,
        x label style={yshift=0.2em},
        ymin=-0.5, ymax=0.5,
        ytick={-0.5,-0.25,0,0.25,0.5},
        xtick={-0.5,-0.25,0,0.25,0.5},
        xticklabels={$-0.5$,$-0.25$,$0$,$0.25$,$0.5$},
        yticklabels={$-0.5$,$-0.25$,$0$,$0.25$,$0.5$},
        ylabel near ticks,
        xlabel={TX \gls{naf}},
        ylabel={RX \gls{naf}},
        y label style={yshift=-2em},
        label style={font=\footnotesize},
        tick label style={font=\footnotesize},
        legend style={font=\footnotesize},
        legend pos = north east,
        legend style=
        	{fill=white, 
        	fill opacity=0.4, 
        	draw opacity=1, 
        	text opacity=1, 
        	nodes={scale=1, transform shape}, 
            /tikz/every even column/.append style={column sep=0.1cm}
        	},
        colorbar,
        point meta min=-60,
        point meta max=0,
        colormap name=jet,
        colorbar style=
            {ylabel={Power [dB]}, 
            at={(1.10, 0)},
            anchor = south,
            width = 0.3cm,
            ytick={-60,-45,-15, 0},
            ylabel style={yshift=0.7cm},
            every axis/.append style=
                {font=\footnotesize}
            }
        ]

      \addplot[forget plot]graphics[
      xmin=-0.5, 
      xmax=0.5, 
      ymin=-0.5, 
      ymax=0.5,
      includegraphics={
        trim=1 0 0 1,
        clip,}
        ] {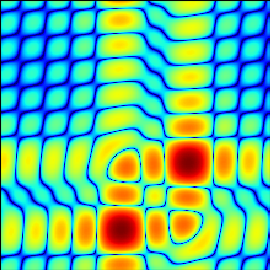};

    \addplot[mark=x, color=white, mark options={scale=2, line width=1.5pt}, mark size=3.0pt] coordinates {(-0.05, -0.35)};
    \addplot[mark=x, color=white, mark options={scale=2, line width=1.5pt}, mark size=3.0pt] coordinates {(0.2, -0.1)};

    \end{axis}

\end{tikzpicture}}
        \vspace{\genvspace}
		}

    \subfloat
  		[Cartesian: Oversampled \\$220\times220$ 2D angles. 
  		\label{fig:intensity_cart_reference_oversampled}]
  		{
		\scalebox{\subscale}{\pgfplotsset{scaled y ticks=false}
\definecolor{steelbluegnu}{RGB}{31,119,180}

\def \globalscale {1}
\begin{tikzpicture}[yscale=\globalscale,xscale=\globalscale]

    \begin{axis}[
        axis on top,
        height=0.315\textwidth,
        width=0.315\textwidth,
        grid=major,
        grid style={solid, black!15},
        enlargelimits=false,
        xmin=-30, xmax=30,
        xlabel near ticks,
        x label style={yshift=0.2em},
        ymin=0, ymax=60,
        ytick={0,15,30,45,60},  
        xtick={-30,-15,0,15,30},
        xticklabels={$-30$,$-15$,$0$,$15$,$30$},
        yticklabels={$0$,$15$,  ,$45$,$60$},  
        ylabel near ticks,
        xlabel={x [m]},
        ylabel={y [m]},
        y label style={yshift=-1.5em},
        label style={font=\footnotesize},
        tick label style={font=\footnotesize},
        legend style={font=\footnotesize},
        legend pos = north east,
        legend style=
        	{fill=white, 
        	fill opacity=0.4, 
        	draw opacity=1, 
        	text opacity=1, 
        	nodes={scale=1, transform shape}, 
            /tikz/every even column/.append style={column sep=0.1cm}
        	},
        colormap name=jet,
        ]

      \addplot[forget plot] graphics[
      xmin=-50, 
      xmax=50, 
      ymin=0, 
      ymax=100,
      includegraphics={
        trim=1 0 0 1,
        clip,}
        ] {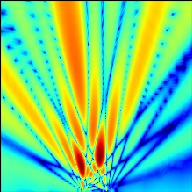};

        \addplot[mark=triangle*, mark options={rotate=180, fill=steelbluegnu}, color=white, mark size=6.5pt] coordinates {
        (-6,0)};
        \addplot[mark=triangle*, mark options={rotate=180, fill=orange}, color=white, mark size=6.5pt] coordinates {
        (6,0)};

    \addplot[mark=x, color=white, mark options={scale=2, line width=1.5pt}, mark size=3.0pt] coordinates {(-7.371, 13.641)};
    \addplot[mark=x, color=white, mark options={scale=2, line width=1.5pt}, mark size=3.0pt] coordinates {(2.176, 18.734)};

    \end{axis}

\end{tikzpicture}}
        \vspace{\genvspace}
		} 
		\hspace{-3.5mm}
  \subfloat
  		[Cartesian: \gls{dft} based samples \\$11\times11$ 2D angles. 
  		\label{fig:intensity_cart_sara_sampled}]
  		{
		\scalebox{\subscale}{\pgfplotsset{scaled y ticks=false}
\definecolor{steelbluegnu}{RGB}{31,119,180}

\def \globalscale {1}
\begin{tikzpicture}[yscale=\globalscale,xscale=\globalscale]

    \begin{axis}[
        axis on top,
        height=0.315\textwidth,
        width=0.315\textwidth,
        grid=major,
        grid style={solid, black!15},
        enlargelimits=false,
        xmin=-30, xmax=30,
        xlabel near ticks,
        x label style={yshift=0.2em},
        ymin=0, ymax=60,
        ytick={0,15,30,45,60},  
        xtick={-30,-15,0,15,30},
        xticklabels={$-30$,$-15$,$0$,$15$,$30$},
        yticklabels={$0$,$15$,  ,$45$,$60$},  
        ylabel near ticks,
        xlabel={x [m]},
        ylabel={y [m]},
        y label style={yshift=-1.5em},
        label style={font=\footnotesize},
        tick label style={font=\footnotesize},
        legend style={font=\footnotesize},
        legend pos = north east,
        legend style=
        	{fill=white, 
        	fill opacity=0.4, 
        	draw opacity=1, 
        	text opacity=1, 
        	nodes={scale=1, transform shape}, 
            /tikz/every even column/.append style={column sep=0.1cm}
        	},
        ]

      \addplot[forget plot]graphics[
      xmin=-50, 
      xmax=50, 
      ymin=0, 
      ymax=100,
      includegraphics={
        trim=1 0 0 1,
        clip,}
        ] {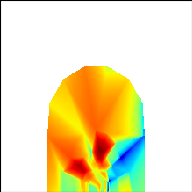};

    \addplot[mark=triangle*, mark options={rotate=180, fill=steelbluegnu}, color=white, mark size=6.5pt] coordinates {
    (-6,0)};
    \addplot[mark=triangle*, mark options={rotate=180, fill=orange}, color=white, mark size=6.5pt] coordinates {
    (6,0)};

    \addplot[mark=x, color=white, mark options={scale=2, line width=1.5pt}, mark size=3.0pt] coordinates {(-7.371, 13.641)};
    \addplot[mark=x, color=white, mark options={scale=2, line width=1.5pt}, mark size=3.0pt] coordinates {(2.176, 18.734)};

    \end{axis}

\end{tikzpicture}}
        \vspace{\genvspace}
		} 
		\hspace{-3.5mm}
  \subfloat
  		[Cartesian: 2D NAF Spline \\$220\times220$ upsampled.
  		\label{fig:intensity_cart_baselineNAF}]
  		{
		\scalebox{\subscale}{\pgfplotsset{scaled y ticks=false}
\definecolor{steelbluegnu}{RGB}{31,119,180}

\def \globalscale {1}
\begin{tikzpicture}[yscale=\globalscale,xscale=\globalscale]

    \begin{axis}[
        axis on top,
        height=0.315\textwidth,
        width=0.315\textwidth,
        grid=major,
        grid style={solid, black!15},
        enlargelimits=false,
        xmin=-30, xmax=30,
        xlabel near ticks,
        x label style={yshift=0.2em},
        ymin=0, ymax=60,
        ytick={0,15,30,45,60},  
        xtick={-30,-15,0,15,30},
        xticklabels={$-30$,$-15$,$0$,$15$,$30$},
        yticklabels={$0$,$15$,  ,$45$,$60$},  
        ylabel near ticks,
        xlabel={x [m]},
        ylabel={y [m]},
        y label style={yshift=-1.5em},
        label style={font=\footnotesize},
        tick label style={font=\footnotesize},
        legend style={font=\footnotesize},
        legend pos = north east,
        legend style=
        	{fill=white, 
        	fill opacity=0.4, 
        	draw opacity=1, 
        	text opacity=1, 
        	nodes={scale=1, transform shape}, 
            /tikz/every even column/.append style={column sep=0.1cm}
        	},
        colormap name=jet,
        ]

      \addplot[forget plot] graphics[
      xmin=-50, 
      xmax=50, 
      ymin=0, 
      ymax=100,
      includegraphics={
        trim=1 0 0 1,
        clip,}
        ] {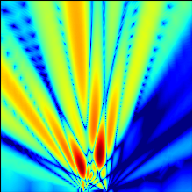};

    \addplot[mark=triangle*, mark options={rotate=180, fill=steelbluegnu}, color=white, mark size=6.5pt] coordinates {
    (-6,0)};
    \addplot[mark=triangle*, mark options={rotate=180, fill=orange}, color=white, mark size=6.5pt] coordinates {
    (6,0)};

    \addplot[mark=x, color=white, mark options={scale=2, line width=1.5pt}, mark size=3.0pt] coordinates {(-7.371, 13.641)};
    \addplot[mark=x, color=white, mark options={scale=2, line width=1.5pt}, mark size=3.0pt] coordinates {(2.176, 18.734)};

    \end{axis}

\end{tikzpicture}}
        \vspace{\genvspace}
		} 
		\hspace{-3.5mm} 
  \subfloat
  		[Cartesian: Bistatic \gls{dft} \\$220\times220$ upsampled.
  		\label{fig:intensity_cart_sara_upsampled}]
  		{
		\scalebox{\subscale}{\pgfplotsset{scaled y ticks=false}
\definecolor{steelbluegnu}{RGB}{31,119,180}

\def \globalscale {1}
\begin{tikzpicture}[yscale=\globalscale,xscale=\globalscale]

    \begin{axis}[
        axis on top,
        height=0.315\textwidth,
        width=0.315\textwidth,
        grid=major,
        grid style={solid, black!15},
        enlargelimits=false,
        xmin=-30, xmax=30,
        xlabel near ticks,
        x label style={yshift=0.2em},
        ymin=0, ymax=60,
        ytick={0,15,30,45,60},  
        xtick={-30,-15,0,15,30},
        xticklabels={$-30$,$-15$,$0$,$15$,$30$},
        yticklabels={$0$,$15$,  ,$45$,$60$},  
        ylabel near ticks,
        xlabel={x [m]},
        ylabel={y [m]},
        y label style={yshift=-1.5em},
        label style={font=\footnotesize},
        tick label style={font=\footnotesize},
        legend style={font=\footnotesize},
        legend pos = north east,
        legend style=
        	{fill=white, 
        	fill opacity=0.4, 
        	draw opacity=1, 
        	text opacity=1, 
        	nodes={scale=1, transform shape}, 
            /tikz/every even column/.append style={column sep=0.1cm}
        	},
        colorbar,
        point meta min=-60,
        point meta max=0,
        colormap name=jet,
        colorbar style=
            {ylabel={Power [dB]}, 
            at={(1.10, 0)},
            anchor = south,
            width = 0.3cm,
            ytick={-60,-45,-15, 0},
            ylabel style={yshift=0.7cm},
            every axis/.append style=
                {font=\footnotesize}
            }
        ]

      \addplot[forget plot]graphics[
      xmin=-50, 
      xmax=50, 
      ymin=0, 
      ymax=100,
      includegraphics={
        trim=1 0 0 1,
        clip,}
        ] {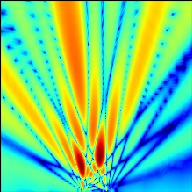};

    \addplot[mark=triangle*, mark options={rotate=180, fill=steelbluegnu}, color=white, mark size=6.5pt] coordinates {
    (-6,0)};
    \addplot[mark=triangle*, mark options={rotate=180, fill=orange}, color=white, mark size=6.5pt] coordinates {
    (6,0)};

    \addplot[mark=x, color=white, mark options={scale=2, line width=1.5pt}, mark size=3.0pt] coordinates {(-7.371, 13.641)};
    \addplot[mark=x, color=white, mark options={scale=2, line width=1.5pt}, mark size=3.0pt] coordinates {(2.176, 18.734)};

    \end{axis}

\end{tikzpicture}}
        \vspace{\genvspace}
		} 
  \caption {Noiseless intensity over 2D bistatic azimuth \gls{naf} (\textit{top row}) and linearly interpolated conversion to Cartesian coordinates (\textit{bottom row}). \textit{From left to right}: Oversampled reference; minimal samples taken according to \gls{dft} principles; \gls{2D} \gls{naf} spline baseline based on minimal samples; and perfect reconstruction of reference achieved by bistatic \gls{dft} based \gls{2D} interpolation also based on minimal samples.}

\label{fig:Intensity_2DBistatic}
\vspace{-4mm}
\end{figure*}

In the qualitative evaluation, different sampling schemes are compared in a noiseless scenario of two fixed ideal point scatterers, as shown in Fig.~\ref{fig:Intensity_2DBistatic}. The top row illustrates the intensity plots over the two azimuth \gls{naf} domains, while the bottom row depicts their translation and interpolation to Cartesian coordinates. \gls{tx} and \gls{rx} positions are indicated by blue and orange triangles, respectively, with the boresight of both arrays pointing in the positive direction of the y-axis. For illustration purposes, the Cartesian values obtained by Eq.~\eqref{eq:Angle2Cartesian} are linearly interpolated to create a gapless grid.

\paragraph{Reference (left | \ref{fig:intensity_reference_oversampled}, \ref{fig:intensity_cart_reference_oversampled})} The reference for the scenario is an oversampling approach with $220$ samples per \gls{naf} domain. The two scatterers are clearly visible, with strong but decreasing sidelobes. 
In the Cartesian plot below, the illumination area of the overlapping \gls{tx} and \gls{rx} beams is observed. At greater distances away from the bistatic baseline, a stretch-out effect of the beams can be noticed. This can be explained by the shrinking delta between the respective angles of the two arrays, converging to monostatic behavior. 

\paragraph{Optimal samples (center left | \ref{fig:intensity_sara_sampled}, \ref{fig:intensity_cart_sara_sampled})} Next, the set of minimal yet optimal samples according to our \gls{dft}-based \gls{2D} azimuth approach is shown. It is evident that with only $11$ samples per \gls{naf} domain, the resolution is low; however, both targets are still clearly observable. The translation to Cartesian coordinates is shown for completeness. Given the limited number of samples, the informative value of the Cartesian plot is reduced. It is important to note that the use of this limited but optimal set of samples is intended in most cases as a preliminary step to interpolation in our methodology.

\paragraph{Baseline (center right | \ref{fig:intensity_baselineNAF}, \ref{fig:intensity_cart_baselineNAF})} The \gls{2D} \gls{naf} spline baseline is based on the previously introduced $11$ \gls{naf} samples per domain, which are interpolated to create $220$ samples per domain. While the two targets remain clearly distinguishable, inconsistencies are observable in both plot types w.r.t. the mainlobes and sidelobes of the oversampled reference.

\paragraph{Our approach (right | \ref{fig:intensity_sara_upsampled}, \ref{fig:intensity_cart_sara_upsampled})}  Finally, the optimal reconstruction capabilities of our \gls{dft} based \gls{2D} azimuth interpolation can be observed. In any noiseless scenario, we get an exact replication of the oversampled reference while drastically reducing the number of samples to 121 compared to 48k in the reference.

\subsection{Monte Carlo Experiments}
As a final validation of our methodology, we conduct a Monte Carlo evaluation across three distinct scenarios with $10$ dB noise power at the \gls{rx}. 
The evaluation results of the scenarios are shown, from left to right, for the probability of missed detection $P^{(\text{MD})}$ in Fig.~\ref{fig:PMD_Eval}, and the \gls{2D} \gls{naf} \gls{rmse} in Fig.~\ref{fig:RMSE_Eval}.
A preliminary distinction of the scenarios can be made by the domain of target placement: scenarios \textit{``Left-Right Sweep''} and \textit{``Near-Far Sweep''} operate in the Cartesian domain of the x,y-plane in meters, while scenario \textit{``\gls{naf} Sweep''} operates in the \gls{naf} domain.

For a detection to be considered valid \wrt the $P^{(\text{MD})}$ or false alarm rate $R^{(\text{FA})}$ its peak has to lie within half the mainlobe width of a true target.
The \gls{2D} \gls{naf} \gls{rmse} is calculated for each true target and the closest detected position using a \gls{cfar} with a desired probability of false alarm $P^{(\text{FA})}$ set to $10^{-3}$.

The $R^{(\text{FA})}$ was evaluated for all scenarios but is not shown below, as \gls{naf} spline and our bistatic \gls{dft} approach almost never detected false targets, with some advantages for the \gls{dft} approach. However, the RAD spline regularly produced values on the order of $10^{-1}$.

Furthermore,the results of the RAD spline over each scenario show that the minimal number of samples is insufficient when applied to the radian domain. Uniformly distributed samples in the radian domain lead to an undersampling of the central region (close to boresight), while oversampling the edges \wrt the angular capabilities of an \gls{ula}~\cite{mandelli2022sampling}. As a result, the angular response of the bistatic setup contains significant distortion, which severely degrades performance. Furthermore, the occurrence of false detections remained at significant levels, necessitating their subsequent rejection during the post-processing stage. In order to streamline the following discourse, the results derived from the RAD spline will be (for the most part) disregarded.

\def\xsweep{Left-Right Sweep}
\paragraph{Scenario ``\xsweep'' (left | \ref{fig:xsweep_pmd}, \ref{fig:xsweep_rmse})} 
Two ideal point scatterers are positioned around a center along the x-axis $\mu$. With fixed y-values their coordinates are given as $[(\mu^{\text{(x)}}\pm3), 16]$, where the x-component of the center is evaluated for ${\mu^{\text{(x)}} \in \mathbb{Z} \mid -20 \le x \le 20}$.


For both metrics, the \gls{2D} \gls{naf} spline baseline shows reasonable performance. The anticipated performance peak of the ``sweet spot'' in the center area between the \gls{tx} and \gls{rx} array is clearly visible. 
Our \gls{dft} based \gls{2D} interpolation approach outperforms both baselines on all performance metrics. There is a $7$\% increase in the width of the $P^{(\mathrm{MD})}$ ``sweet spot'' area compared to \gls{naf} spline. 

\def\ysweep{Near-Far Sweep}
\paragraph{Scenario ``\ysweep'' (center | \ref{fig:ysweep_pmd}, \ref{fig:ysweep_rmse})} 
Two ideal point scatterers are positioned at $[\pm3), 16]$, as such with identical y-values and fixed offsets in x direction. The y-value of the targets is evaluated for ${y \in \mathbb{N} \mid 5 \le y \le 45}$.

At approximately $\mathrm{y}=30~\text{m}$ a significant change in performance becomes evident. The systems's angular resolution capabilities in combination with the corresponding interpolation approach are exhausted w.r.t. the distance of operation. Consequently, the two targets merge into a single peak around the $30$~m mark, leading to a general  performance deterioration. For RAD spline, it has been observed that occasionally both targets are missed due the merging of the peaks in combination with the distortion effects resulting from operating in radian.

The \gls{naf} spline experiences the merging of the targets at a distance of $28$~m. Notably,  our bistatic \gls{dft} approach extends this distance to $32$~m, thereby increasing the reliable operations distance for target discrimination by 14\%.



\def\nafsweep{\gls{naf} Sweep}
\paragraph{Scenario ``\nafsweep'' (right | \ref{fig:nafsweep_pmd}, \ref{fig:nafsweep_rmse})} 
In the last scenario, the domain of operation changes to the \gls{2D} \gls{naf}. The two ideal point scatterers are positioned with a fixed offset $\Delta d^{(\text{P})}$. This scenario provides a more unbiased evaluation of the angular resolution capabilities of the system by avoiding the geometric dependencies of the previous Cartesian scenarios. The evaluation sweeps the offset of the two scatterers with ${\Delta d^{(\text{P})} \in \mathbb{R} \mid 0.06 \le \Delta d^{(\text{P})} \le 0.16}$.

\begin{remark}
It is important to note that certain combinations of TX and RX angles in \gls{naf} do not have a physical representation. However, in context of the reconstruction of the angular response they are still required. 
\end{remark}

As previously indicated, the distortion effects of operating in radian, particularly towards the edges, degrade the performance of the RAD spline approach, thus causing its failure in this scenario.

The results of the \nafsweep are categorized into three segments. 

In the first segment, defined for the \gls{2D} \gls{naf} offsets $\Delta d^{(\text{P})}$ below $0.11$, the proximity of the two point targets has led to their merging, which occurs when the resolution capabilities of the system are exceeded. Thus, each approach can only resolve a single target. 

The second segment covers the gradual separation of the merged targets and varies in its length depending on the approach. For the \gls{naf} spline this segment spans for $0.11 \leq \Delta d^{(\text{P})} \leq 0.15$, but for our \gls{dft} based interpolation it only spans for $0.11 \leq \Delta d^{(\text{P})} \leq 0.135$. Our approach gradually outperforms the spline baselines \wrt the $P^{(\text{MD})}$ and \gls{2D} \gls{naf} \gls{rmse}.

The final segment covers the reliable detection of both targets by any \gls{naf}-based methodologies. For our \gls{dft} based approach, this segment starts with the lowest values for $\Delta d^{(\text{P})} > 0.135$, demonstrating superior performance in terms of any performance metric when compared to the baselines.

In summary, irrespective of the scenario or metric, the bistatic \gls{dft} approach has been demonstrated to be either superior to or equivalent to the investigated baseline solutions. Consequently, it provides the most reliable performance with regard to the angular discrimination of multiple targets.

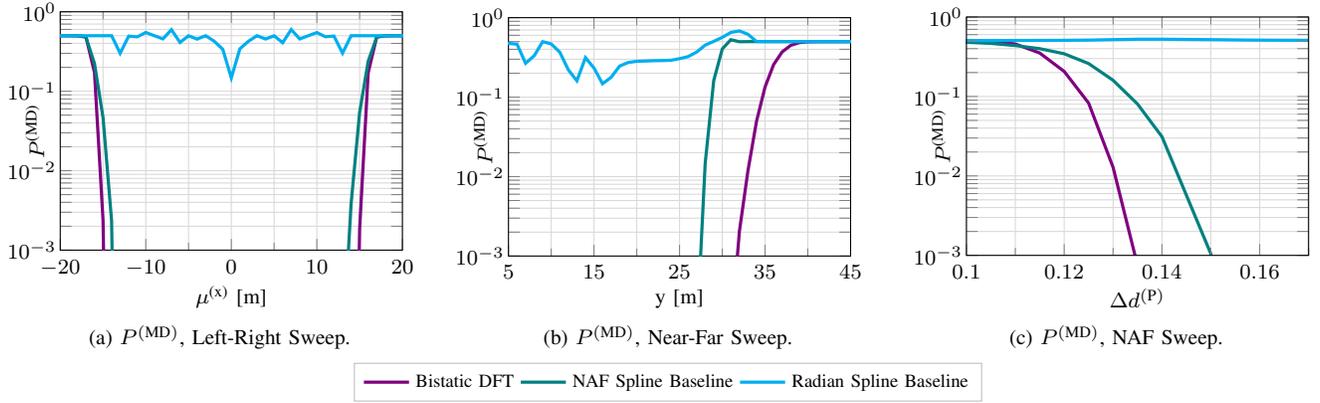
\begin{figure*}[ht]
        \def\scale{.51}  
        \def\cappmd{\(P^{(\text{MD})}\)}
        \def\capfar{\(R^{(\text{FA})}\)}
        \def\cappp{\(P^{(\text{P})}\)}
        \def\capfone{\(F_1\)}
        \def\caprmse{2D NAF RMSE}
        \def\capysweep{, y-sweep}
        \def\capnafsweep{, NAF-sweep}
        \def\subfigwidth{0.329} 
        \def\genvspace{-1.0mm}
        \def\genhspace{-5.0mm}  
        \def\figwidth{10.5cm}
        \def\figheight{7.8cm}
        
        \centering

        \def\capsweep{, \xsweep.}
        \def\dirfig{Cartesian/X_sweep_same_y}
        \begin{subfigure}[t]{\subfigwidth\textwidth}  
            \centering
            \begin{tikzpicture}

\begin{semilogyaxis}[
	xlabel= {$\Delta d^{(\text{P})}$},
    ylabel={$P^\text{(MD)}$},
    ylabel style={font=\footnotesize,at={(axis description cs:.-0.02,.5)},rotate=0,anchor=south},
    x label style={yshift=0.3em},
    ytick={0.00001, 0.00002, 0.00003, 0.00004, 0.00005, 0.00006, 0.00007, 0.00008, 0.00009, 0.0001, 0.0002, 0.0003, 0.0004, 0.0005, 0.0006, 0.0007, 0.0008, 0.0009, 0.001, 0.002, 0.003, 0.004, 0.005, 0.006, 0.007, 0.008, 0.009, 0.01, 0.02, 0.03, 0.04, 0.05, 0.06, 0.07, 0.08, 0.09, 0.1, 0.2, 0.3, 0.4, 0.5, 0.6, 0.7, 0.8, 0.9, 1, 2, 3, 4, 5, 6},
    yticklabels={$10^{-5}$, , , , , , , , ,$10^{-4}$, , , , , , , , ,$10^{-3}$, , , , , , , , , $10^{-2}$ , , , , , , , , ,$10^{-1}$, , , , , , , , , $10^{0}$},
    xlabel style={font=\footnotesize},
    yticklabel style={font=\footnotesize,xshift=2pt},  
    xticklabel style={
        /pgf/number format/precision=4,
        /pgf/number format/fixed, 
        font=\footnotesize},
    ymin=1e-3,
    ymax=1e-0,
    xmin=0.10,
    xmax=0.17,
    scaled x ticks = false,
    minor tick num=1,
    yminorticks = true,
    enlargelimits = false,
    grid = both,
    grid style={solid, black!15},
    legend cell align={left},
    legend columns = 3,
    legend style={
      fill opacity=1.0,
      draw opacity=1,
      text opacity=1,
      at={(0.999,0.999)},  
      anchor=north east,
      draw=lightgray,
      font=\scriptsize
    },
    every axis plot/.append style={very thick},
    mark repeat={2},
    log origin y=infty,
    set layers=Bowpark,
    legend to name = common_legend,
    width=\figwidth,
    height=\figheight,
    scale = \scale
]

\addplot[very thick, color=violet, solid, draw=none] coordinates {(0, 1)}; \addlegendentry{Bistatic \gls{dft}}
\addplot[very thick, color=teal, solid, draw=none] coordinates {(0, 1)}; \addlegendentry{\gls{naf} Spline Baseline}
\addplot[very thick, color=cyan, solid, draw=none] coordinates {(0, 1)}; \addlegendentry{Radian Spline Baseline}


\def\dirstr{"NAF_sweep"}
\def\evalstr{"prob_missed_detect"}

\addplot [very thick, violet, solid] plot table[x index=0, y index=1] {Data/\dirstr/\evalstr.txt};
\addplot [very thick, teal, solid] plot table[x index=0, y index=1] {Data/\dirstr/\evalstr_baseline.txt};
\addplot [very thick, cyan, solid] plot table[x index=0, y index=1] {Data/\dirstr/\evalstr_baseline_rad.txt};


\end{semilogyaxis}

\end{tikzpicture}
            \vspace*{\genvspace}
            \caption{\cappmd\capsweep}
            \label{fig:xsweep_pmd}
        \end{subfigure}
        \hspace*{\genhspace}
        \def\capsweep{, \ysweep.}
        \def\dirfig{Cartesian/Y_sweep_same_y}
        \begin{subfigure}[t]{\subfigwidth\textwidth}  
            \centering
            \begin{tikzpicture}

\begin{semilogyaxis}[
	xlabel= {$\Delta d^{(\text{P})}$},
    ylabel={$P^\text{(MD)}$},
    ylabel style={font=\footnotesize,at={(axis description cs:.-0.02,.5)},rotate=0,anchor=south},
    x label style={yshift=0.3em},
    ytick={0.00001, 0.00002, 0.00003, 0.00004, 0.00005, 0.00006, 0.00007, 0.00008, 0.00009, 0.0001, 0.0002, 0.0003, 0.0004, 0.0005, 0.0006, 0.0007, 0.0008, 0.0009, 0.001, 0.002, 0.003, 0.004, 0.005, 0.006, 0.007, 0.008, 0.009, 0.01, 0.02, 0.03, 0.04, 0.05, 0.06, 0.07, 0.08, 0.09, 0.1, 0.2, 0.3, 0.4, 0.5, 0.6, 0.7, 0.8, 0.9, 1, 2, 3, 4, 5, 6},
    yticklabels={$10^{-5}$, , , , , , , , ,$10^{-4}$, , , , , , , , ,$10^{-3}$, , , , , , , , , $10^{-2}$ , , , , , , , , ,$10^{-1}$, , , , , , , , , $10^{0}$},
    xlabel style={font=\footnotesize},
    yticklabel style={font=\footnotesize,xshift=2pt},  
    xticklabel style={
        /pgf/number format/precision=4,
        /pgf/number format/fixed, 
        font=\footnotesize},
    ymin=1e-3,
    ymax=1e-0,
    xmin=0.10,
    xmax=0.17,
    scaled x ticks = false,
    minor tick num=1,
    yminorticks = true,
    enlargelimits = false,
    grid = both,
    grid style={solid, black!15},
    legend cell align={left},
    legend columns = 3,
    legend style={
      fill opacity=1.0,
      draw opacity=1,
      text opacity=1,
      at={(0.999,0.999)},  
      anchor=north east,
      draw=lightgray,
      font=\scriptsize
    },
    every axis plot/.append style={very thick},
    mark repeat={2},
    log origin y=infty,
    set layers=Bowpark,
    legend to name = common_legend,
    width=\figwidth,
    height=\figheight,
    scale = \scale
]

\addplot[very thick, color=violet, solid, draw=none] coordinates {(0, 1)}; \addlegendentry{Bistatic \gls{dft}}
\addplot[very thick, color=teal, solid, draw=none] coordinates {(0, 1)}; \addlegendentry{\gls{naf} Spline Baseline}
\addplot[very thick, color=cyan, solid, draw=none] coordinates {(0, 1)}; \addlegendentry{Radian Spline Baseline}


\def\dirstr{"NAF_sweep"}
\def\evalstr{"prob_missed_detect"}

\addplot [very thick, violet, solid] plot table[x index=0, y index=1] {Data/\dirstr/\evalstr.txt};
\addplot [very thick, teal, solid] plot table[x index=0, y index=1] {Data/\dirstr/\evalstr_baseline.txt};
\addplot [very thick, cyan, solid] plot table[x index=0, y index=1] {Data/\dirstr/\evalstr_baseline_rad.txt};


\end{semilogyaxis}

\end{tikzpicture}
            \vspace*{\genvspace}
            \caption{\cappmd\capsweep}
            \label{fig:ysweep_pmd}
        \end{subfigure}
        \hspace*{\genhspace}
        \def\capsweep{, \nafsweep.}
        \def\dirfig{NAF}
        \begin{subfigure}[t]{\subfigwidth\textwidth}  
            \centering
            \begin{tikzpicture}

\begin{semilogyaxis}[
	xlabel= {$\Delta d^{(\text{P})}$},
    ylabel={$P^\text{(MD)}$},
    ylabel style={font=\footnotesize,at={(axis description cs:.-0.02,.5)},rotate=0,anchor=south},
    x label style={yshift=0.3em},
    ytick={0.00001, 0.00002, 0.00003, 0.00004, 0.00005, 0.00006, 0.00007, 0.00008, 0.00009, 0.0001, 0.0002, 0.0003, 0.0004, 0.0005, 0.0006, 0.0007, 0.0008, 0.0009, 0.001, 0.002, 0.003, 0.004, 0.005, 0.006, 0.007, 0.008, 0.009, 0.01, 0.02, 0.03, 0.04, 0.05, 0.06, 0.07, 0.08, 0.09, 0.1, 0.2, 0.3, 0.4, 0.5, 0.6, 0.7, 0.8, 0.9, 1, 2, 3, 4, 5, 6},
    yticklabels={$10^{-5}$, , , , , , , , ,$10^{-4}$, , , , , , , , ,$10^{-3}$, , , , , , , , , $10^{-2}$ , , , , , , , , ,$10^{-1}$, , , , , , , , , $10^{0}$},
    xlabel style={font=\footnotesize},
    yticklabel style={font=\footnotesize,xshift=2pt},  
    xticklabel style={
        /pgf/number format/precision=4,
        /pgf/number format/fixed, 
        font=\footnotesize},
    ymin=1e-3,
    ymax=1e-0,
    xmin=0.10,
    xmax=0.17,
    scaled x ticks = false,
    minor tick num=1,
    yminorticks = true,
    enlargelimits = false,
    grid = both,
    grid style={solid, black!15},
    legend cell align={left},
    legend columns = 3,
    legend style={
      fill opacity=1.0,
      draw opacity=1,
      text opacity=1,
      at={(0.999,0.999)},  
      anchor=north east,
      draw=lightgray,
      font=\scriptsize
    },
    every axis plot/.append style={very thick},
    mark repeat={2},
    log origin y=infty,
    set layers=Bowpark,
    legend to name = common_legend,
    width=\figwidth,
    height=\figheight,
    scale = \scale
]

\addplot[very thick, color=violet, solid, draw=none] coordinates {(0, 1)}; \addlegendentry{Bistatic \gls{dft}}
\addplot[very thick, color=teal, solid, draw=none] coordinates {(0, 1)}; \addlegendentry{\gls{naf} Spline Baseline}
\addplot[very thick, color=cyan, solid, draw=none] coordinates {(0, 1)}; \addlegendentry{Radian Spline Baseline}


\def\dirstr{"NAF_sweep"}
\def\evalstr{"prob_missed_detect"}

\addplot [very thick, violet, solid] plot table[x index=0, y index=1] {Data/\dirstr/\evalstr.txt};
\addplot [very thick, teal, solid] plot table[x index=0, y index=1] {Data/\dirstr/\evalstr_baseline.txt};
\addplot [very thick, cyan, solid] plot table[x index=0, y index=1] {Data/\dirstr/\evalstr_baseline_rad.txt};


\end{semilogyaxis}

\end{tikzpicture}
            \vspace*{\genvspace}
            \caption{\cappmd\capsweep}
            \label{fig:nafsweep_pmd}
            \vspace*{1.5mm}
        \end{subfigure}
        
        \centerline{{\hypersetup{hidelinks}\ref{common_legend}}}
        \caption{Probability of missed detection \cappmd evaluations for three bistatic azimuth-only scenarios with varying placement of two ideal point scatterers and $\sigma^2=10$ dB. Scenario~(A)~(\textit{left}): fixed x-axis offset of $6$ m, sweeping from left to right. Scenario~(B)~(\textit{center}): fixed x-axis offset of $5$ m, sweeping from near to far. Scenario~(C)~(\textit{right}): sweeping offset in the \gls{2D} azimuth \gls{naf} domain.} 
\label{fig:PMD_Eval}
\vspace{-4mm}
\end{figure*}

\begin{figure*}[ht]
        \def\scale{.51}  
        \def\cappmd{\(P^{(\text{MD})}\)}
        \def\capfar{\(R^{(\text{FA})}\)}
        \def\cappp{\(P^{(\text{P})}\)}
        \def\capfone{\(F_1\)}
        \def\caprmse{2D NAF RMSE}
        \def\capysweep{, y-sweep}
        \def\capnafsweep{, NAF-sweep}
        \def\subfigwidth{0.329} 
        \def\genvspace{-1.0mm}
        \def\genhspace{-5.0mm}  
        \def\figwidth{10.5cm}
        \def\figheight{7.8cm}
        
        \centering

        \def\capsweep{, \xsweep.}
        \def\dirfig{Cartesian/X_sweep_same_y}
        \begin{subfigure}[t]{\subfigwidth\textwidth}
            \centering
            \begin{tikzpicture}

\begin{semilogyaxis}[
    xlabel= {$\Delta d^{(\text{P})}$},
    ylabel={$\text{2D NAF RMSE}$},
    ylabel style={font=\footnotesize,at={(axis description cs:.-0.02,.5)},rotate=0,anchor=south},
    x label style={yshift=0.3em},
    ytick={0.00001, 0.00002, 0.00003, 0.00004, 0.00005, 0.00006, 0.00007, 0.00008, 0.00009, 0.0001, 0.0002, 0.0003, 0.0004, 0.0005, 0.0006, 0.0007, 0.0008, 0.0009, 0.001, 0.002, 0.003, 0.004, 0.005, 0.006, 0.007, 0.008, 0.009, 0.01, 0.02, 0.03, 0.04, 0.05, 0.06, 0.07, 0.08, 0.09, 0.1, 0.2, 0.3, 0.4, 0.5, 0.6, 0.7, 0.8, 0.9, 1, 2, 3, 4, 5, 6},
    yticklabels={$10^{-5}$, , , , , , , , ,$10^{-4}$, , , , , , , , ,$10^{-3}$, , , , , , , , , $\mathbf{10^{-2}}$ , , , , , , , , ,  , , , , , , , , , $10^{0}$},
    xlabel style={font=\footnotesize},
    yticklabel style={font=\footnotesize,xshift=2pt},  
    xticklabel style={
        /pgf/number format/precision=4,
        /pgf/number format/fixed, 
        font=\footnotesize},
    ymin=1e-2,
    ymax=1e-0,
    xmin=0.10,
    xmax=0.17,
    scaled x ticks = false,
    minor tick num=1,
    yminorticks = true,
    enlargelimits = false,
    grid = both,
    grid style={solid, black!15},
    legend cell align={left},
    legend columns = 4,
    legend style={
      fill opacity=1.0,
      draw opacity=1,
      text opacity=1,
      at={(0.999,0.999)},  
      anchor=north east,
      draw=lightgray,
      font=\scriptsize
    },
    every axis plot/.append style={very thick},
    mark repeat={2},
    log origin y=infty,
    set layers=Bowpark,
    width=\figwidth,
    height=\figheight,
    scale = \scale
]


\def\dirstr{"NAF_sweep"}
\def\evalstr{"rmse"}

\addplot [very thick, violet, solid] plot table[x index=0, y index=1] {Data/\dirstr/\evalstr.txt};
\addplot [very thick, teal, solid] plot table[x index=0, y index=1] {Data/\dirstr/\evalstr_baseline.txt};
\addplot [very thick, cyan, solid] plot table[x index=0, y index=1] {Data/\dirstr/\evalstr_baseline_rad.txt};

\end{semilogyaxis}

\end{tikzpicture}
            \vspace*{\genvspace}
            \caption{\caprmse\capsweep}
            \label{fig:xsweep_rmse}
        \end{subfigure}
        \hspace*{\genhspace}
        \def\capsweep{, \ysweep.}
        \def\dirfig{Cartesian/Y_sweep_same_y}
        \begin{subfigure}[t]{\subfigwidth\textwidth}
            \centering
            \begin{tikzpicture}

\begin{semilogyaxis}[
    xlabel= {$\Delta d^{(\text{P})}$},
    ylabel={$\text{2D NAF RMSE}$},
    ylabel style={font=\footnotesize,at={(axis description cs:.-0.02,.5)},rotate=0,anchor=south},
    x label style={yshift=0.3em},
    ytick={0.00001, 0.00002, 0.00003, 0.00004, 0.00005, 0.00006, 0.00007, 0.00008, 0.00009, 0.0001, 0.0002, 0.0003, 0.0004, 0.0005, 0.0006, 0.0007, 0.0008, 0.0009, 0.001, 0.002, 0.003, 0.004, 0.005, 0.006, 0.007, 0.008, 0.009, 0.01, 0.02, 0.03, 0.04, 0.05, 0.06, 0.07, 0.08, 0.09, 0.1, 0.2, 0.3, 0.4, 0.5, 0.6, 0.7, 0.8, 0.9, 1, 2, 3, 4, 5, 6},
    yticklabels={$10^{-5}$, , , , , , , , ,$10^{-4}$, , , , , , , , ,$10^{-3}$, , , , , , , , , $\mathbf{10^{-2}}$ , , , , , , , , ,  , , , , , , , , , $10^{0}$},
    xlabel style={font=\footnotesize},
    yticklabel style={font=\footnotesize,xshift=2pt},  
    xticklabel style={
        /pgf/number format/precision=4,
        /pgf/number format/fixed, 
        font=\footnotesize},
    ymin=1e-2,
    ymax=1e-0,
    xmin=0.10,
    xmax=0.17,
    scaled x ticks = false,
    minor tick num=1,
    yminorticks = true,
    enlargelimits = false,
    grid = both,
    grid style={solid, black!15},
    legend cell align={left},
    legend columns = 4,
    legend style={
      fill opacity=1.0,
      draw opacity=1,
      text opacity=1,
      at={(0.999,0.999)},  
      anchor=north east,
      draw=lightgray,
      font=\scriptsize
    },
    every axis plot/.append style={very thick},
    mark repeat={2},
    log origin y=infty,
    set layers=Bowpark,
    width=\figwidth,
    height=\figheight,
    scale = \scale
]


\def\dirstr{"NAF_sweep"}
\def\evalstr{"rmse"}

\addplot [very thick, violet, solid] plot table[x index=0, y index=1] {Data/\dirstr/\evalstr.txt};
\addplot [very thick, teal, solid] plot table[x index=0, y index=1] {Data/\dirstr/\evalstr_baseline.txt};
\addplot [very thick, cyan, solid] plot table[x index=0, y index=1] {Data/\dirstr/\evalstr_baseline_rad.txt};

\end{semilogyaxis}

\end{tikzpicture}
            \vspace*{\genvspace}
            \caption{\caprmse\capsweep}
            \label{fig:ysweep_rmse}
        \end{subfigure}
        \hspace*{\genhspace}
        \def\capsweep{, \nafsweep.}
        \def\dirfig{NAF}
        \begin{subfigure}[t]{\subfigwidth\textwidth}
            \centering
            \begin{tikzpicture}

\begin{semilogyaxis}[
    xlabel= {$\Delta d^{(\text{P})}$},
    ylabel={$\text{2D NAF RMSE}$},
    ylabel style={font=\footnotesize,at={(axis description cs:.-0.02,.5)},rotate=0,anchor=south},
    x label style={yshift=0.3em},
    ytick={0.00001, 0.00002, 0.00003, 0.00004, 0.00005, 0.00006, 0.00007, 0.00008, 0.00009, 0.0001, 0.0002, 0.0003, 0.0004, 0.0005, 0.0006, 0.0007, 0.0008, 0.0009, 0.001, 0.002, 0.003, 0.004, 0.005, 0.006, 0.007, 0.008, 0.009, 0.01, 0.02, 0.03, 0.04, 0.05, 0.06, 0.07, 0.08, 0.09, 0.1, 0.2, 0.3, 0.4, 0.5, 0.6, 0.7, 0.8, 0.9, 1, 2, 3, 4, 5, 6},
    yticklabels={$10^{-5}$, , , , , , , , ,$10^{-4}$, , , , , , , , ,$10^{-3}$, , , , , , , , , $\mathbf{10^{-2}}$ , , , , , , , , ,  , , , , , , , , , $10^{0}$},
    xlabel style={font=\footnotesize},
    yticklabel style={font=\footnotesize,xshift=2pt},  
    xticklabel style={
        /pgf/number format/precision=4,
        /pgf/number format/fixed, 
        font=\footnotesize},
    ymin=1e-2,
    ymax=1e-0,
    xmin=0.10,
    xmax=0.17,
    scaled x ticks = false,
    minor tick num=1,
    yminorticks = true,
    enlargelimits = false,
    grid = both,
    grid style={solid, black!15},
    legend cell align={left},
    legend columns = 4,
    legend style={
      fill opacity=1.0,
      draw opacity=1,
      text opacity=1,
      at={(0.999,0.999)},  
      anchor=north east,
      draw=lightgray,
      font=\scriptsize
    },
    every axis plot/.append style={very thick},
    mark repeat={2},
    log origin y=infty,
    set layers=Bowpark,
    width=\figwidth,
    height=\figheight,
    scale = \scale
]


\def\dirstr{"NAF_sweep"}
\def\evalstr{"rmse"}

\addplot [very thick, violet, solid] plot table[x index=0, y index=1] {Data/\dirstr/\evalstr.txt};
\addplot [very thick, teal, solid] plot table[x index=0, y index=1] {Data/\dirstr/\evalstr_baseline.txt};
\addplot [very thick, cyan, solid] plot table[x index=0, y index=1] {Data/\dirstr/\evalstr_baseline_rad.txt};

\end{semilogyaxis}

\end{tikzpicture}
            \vspace*{\genvspace}
            \caption{\caprmse\capsweep}
            \label{fig:nafsweep_rmse}
            \vspace*{1.5mm}
        \end{subfigure}
        
        \centerline{{\hypersetup{hidelinks}\ref{common_legend}}}
        \caption{\gls{2D} \gls{naf} \gls{rmse} evaluations for three bistatic azimuth-only scenarios with varying placement of two ideal point scatterers and $\sigma^2=10$ dB. Scenario~(A)~(\textit{left}): fixed x-axis offset of $6$ m, sweeping from left to right. Scenario~(B)~(\textit{center}): fixed x-axis offset of $5$ m, sweeping from near to far. Scenario~(C)~(\textit{right}): sweeping offset in the \gls{2D} azimuth \gls{naf} domain.} 
\label{fig:RMSE_Eval}
\vspace{-4mm}
\end{figure*}
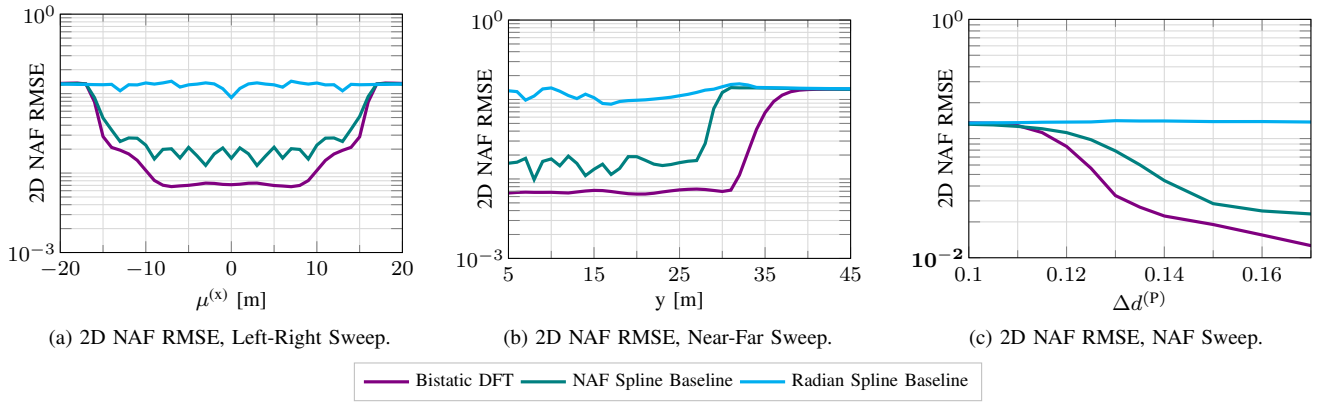
\section{Conclusion}


In this work, we present our solution to efficiently sample angular domains in bistatic \gls{isac} deployments.

Extending the considerations done for monostatic setups in~\cite{mandelli2022sampling}, we determine the minimal number of angular sampling points for a given number of array elements, thus, ensuring a methodology with minimal overhead. This is possible by exploiting the Fourier duality of the angular sampling domain and the array element position domain.
Further, we demonstrated that interpolation with Dirichlet kernels can also be applied to \gls{2D} azimuth-only scenarios enabling perfect reconstruction capabilities of the angular response. 

In our simulation study we demonstrate superior performance on multiple metrics (probability of missed detection, false alarm rate, and \gls{2D} \gls{rmse}) when compared to different variations of a \gls{2D} spline interpolation baseline.

In future studies, we plan to extend the proposed angular sampling and interpolation methods to include the elevation of the TX and RX arrays.

\appendices


\section*{Acknowledgment}
The authors would like to thank Maximilian Bauhofer for his support during the development of this work.

This work was developed within the KOMSENS-6G project, partly funded by the German Ministry of Education and Research under grant 16KISK112K.

\balance
\else

\appendices



\fi

\bibliographystyle{IEEEtran}
\bibliography{bistatic2D_main}

\begin{thebibliography}{10}
\providecommand{\url}[1]{#1}
\csname url@samestyle\endcsname
\providecommand{\newblock}{\relax}
\providecommand{\bibinfo}[2]{#2}
\providecommand{\BIBentrySTDinterwordspacing}{\spaceskip=0pt\relax}
\providecommand{\BIBentryALTinterwordstretchfactor}{4}
\providecommand{\BIBentryALTinterwordspacing}{\spaceskip=\fontdimen2\font plus
\BIBentryALTinterwordstretchfactor\fontdimen3\font minus \fontdimen4\font\relax}
\providecommand{\BIBforeignlanguage}[2]{{%
\expandafter\ifx\csname l@#1\endcsname\relax
\typeout{** WARNING: IEEEtran.bst: No hyphenation pattern has been}%
\typeout{** loaded for the language `#1'. Using the pattern for}%
\typeout{** the default language instead.}%
\else
\language=\csname l@#1\endcsname
\fi
#2}}
\providecommand{\BIBdecl}{\relax}
\BIBdecl

\bibitem{3GPP_TR_22837}
{3GPP}, ``{Study on Integrated Sensing and Communication},'' 3rd Generation Partnership Project (3GPP), Technical Report TR 22.837, June 2023.

\bibitem{wild2021joint}
T.~Wild, V.~Braun, and H.~Viswanathan, ``{Joint Design of Communication and Sensing for Beyond 5G and 6G Systems},'' \emph{IEEE Access}, vol.~9, pp. 30\,845--30\,857, Feb. 2021.

\bibitem{felixAntennaArrayDesign2024a}
A.~Felix, S.~Mandelli, M.~Henninger, and S.~Ten~Brink, ``Antenna {{Array Design}} for {{Monostatic ISAC}},'' in \emph{2024 {{IEEE}} 25th {{International Workshop}} on {{Signal Processing Advances}} in {{Wireless Communications}} ({{SPAWC}})}, pp. 721--725.

\bibitem{wild2023integrated}
T.~Wild, A.~Grudnitsky, S.~Mandelli, M.~Henninger, J.~Guan, and F.~Schaich, ``{6G Integrated Sensing and Communication: From Vision to Realization},'' in \emph{2023 20th European Radar Conference (EuRAD)}, Sep. 2023, pp. 355--358.

\bibitem{richardsFundamentalsRadarSignal2014}
M.~Richards, \emph{Fundamentals of {{Radar Signal Processing}}, {{Second Edition}}}, 2nd~ed.\hskip 1em plus 0.5em minus 0.4em\relax McGraw Hill.

\bibitem{skolnikIntroductionRadarSystems2002}
M.~Skolnik, \emph{Introduction to {{Radar Systems}}}, 3rd~ed.\hskip 1em plus 0.5em minus 0.4em\relax McGraw-Hill Education, December 2002.

\bibitem{sitCharacterizingEvaporationDucts2019}
H.~Sit and C.~J. Earls, ``Characterizing {{Evaporation Ducts Within}} the {{Marine Atmospheric Boundary Layer Using Artificial Neural Networks}},'' vol.~54, no.~12, pp. 1181--1191.

\bibitem{panFeasibilityStudyPassive2019}
J.~Pan, P.~Hu, Q.~Zhu, Q.~Bao, and Z.~Chen, ``Feasibility {{Study}} of {{Passive Bistatic Radar Based}} on {{Phased Array Radar Signals}},'' vol.~8, no.~7, p. 728, July 2019.

\bibitem{rajamaki2019analog}
R.~Rajam{\"a}ki, S.~P. Chepuri, and V.~Koivunen, ``Analog beamforming for active imaging using sparse arrays,'' in \emph{2019 53rd Asilomar Conference on Signals, Systems, and Computers}, 2019, pp. 1202--1206.

\bibitem{rajamaki2020hybrid}
------, ``Hybrid beamforming for active sensing using sparse arrays,'' \emph{IEEE Transactions on Signal Processing}, vol.~68, pp. 6402--6417, Oct. 2020.

\bibitem{mandelli2022sampling}
S.~Mandelli, M.~Henninger, and J.~Du, ``Sampling and reconstructing angular domains with uniform arrays,'' \emph{IEEE Transactions on Wireless Communications}, Nov. 2022.

\bibitem{hoctor1990unifying}
R.~T. Hoctor and S.~A. Kassam, ``The unifying role of the coarray in aperture synthesis for coherent and incoherent imaging,'' \emph{Proceedings of the IEEE}, vol.~78, no.~4, pp. 735--752, Apr. 1990.

\bibitem{virtanenSciPyFundamentalAlgorithms2020}
P.~Virtanen \emph{et~al.}, ``{{SciPy}} 1.0: Fundamental algorithms for scientific computing in {{Python}},'' \emph{Nature Methods}, vol.~17, no.~3, pp. 261--272.

\end{thebibliography}

\end{document}